\documentclass[fleqn,usenatbib]{mnras}

\usepackage{newtxtext,newtxmath}

\usepackage[T1]{fontenc}
\usepackage{ae,aecompl}

\usepackage{graphicx}	
\usepackage{amsmath}	
\usepackage{amssymb}	

\usepackage[markup=underlined]{changes}	
\usepackage{color}

\definechangesauthor[color=red]{AB}
\definechangesauthor[color=blue]{DM}
\definechangesauthor[color=green]{AL}

\title[Andromeda and Beyond]{Through Andromeda and Beyond: AGN Optical Transient `Sharov21' Revisited}

\author[A. Bruce et al.]{
A. Bruce,\thanks{E-mail: alb@roe.ac.uk}
A. Lawrence,
D. McLeod,
Nicholas P. Ross,
\\
Institute for Astronomy, SUPA (Scottish Universities Physics Alliance), University of Edinburgh, Royal Observatory, Blackford Hill, Edinburgh EH9 3HJ, Edinburgh, UK\\
}

\date{Accepted XXX. Received YYY; in original form ZZZ}

\pubyear{2019}

\begin{document}
\label{firstpage}
\pagerange{\pageref{firstpage}--\pageref{lastpage}}
\maketitle

\begin{abstract}
We revisit a notable AGN known as `Sharov21', seen to undergo a dramatic outburst in 1992, brightening by a factor of thirty over a period of approximately one year. A simple microlensing model fit to the event lightcurve provides a constraint on the distance of the lensing object which is consistent with the distance to M31, strongly suggesting that this is the correct explanation. Archival XMM/Hubble/Spitzer data show that this AGN can be considered an otherwise unremarkable type-I AGN. Our analysis of the expected rate of background AGN being microlensed by a factor of two or more due to stellar-mass objects in M31 shows that events of this nature should only occur on average every half century. It is thus perhaps surprising that we have uncovered evidence for two more events that are qualitatively similar. A systematic search for new and archival events, with follow-up spectroscopy, is thus warranted.
\end{abstract}




\section{Introduction}

The variability exhibited by active galactic nuclei (AGN) has been studied extensively but the physical mechanisms responsible for these variations remain difficult to pin down. On timescales greater than a few days, variations are typically on the order of 0.2 mag and the variability amplitude tends to increase with frequency. While the exact cause of the intrinsic variations remains unknown, the variability allows a probe of the inner structure of AGN, namely the radial profile of both the broad line region (BLR) and accretion disc (e.g.: \citet{Peterson2004,MacLeod2012,Homayouni2019} and review by \citet{Lawrence2016b}). Variability of an extrinsic nature, such as the ongoing microlensing `flickering' seen in multiply-imaged AGN, also allows us to probe the structure of the inner regions \citep{Morgan2010, Motta2017}.

Rarely, AGN are seen to undergo longer-lived outbursts which can rise an order of magnitude above quiescence. The AGN known as `Sharov21' is one such extreme example. First discovered by \citet{Nedialkov1996} and reported in \citet{Sharov1998}, this object was interpreted as a nova residing in M31. A later spectroscopic confirmation and detailed analysis from \citet{Meusinger2010} showed that Sharov21 was in fact an AGN at the much greater redshift of $z=2.109$. The outburst seen in this object was on the order of one year in duration and rose to more than three magnitudes above the background, with no other outbursts evident in the decades-long lightcurve. \citet{Meusinger2010} explore two explanations for this outburst: 1) that it was caused by the tidal disruption of a $\sim$10\,M$_\odot$ star in close proximity to an extant accretion disc and 2) that it was caused by a microlensing event due to a stellar-mass lens in M31. In their work the former of these two scenarios is favoured, in part due to the low probability of a microlensing event occurring.

In this paper, we wish to re-examine the microlensing hypothesis. Recent observations have suggested that a number of long-lived AGN transients can be explained as the result of isolated microlensing events \citep{Lawrence2016a,Bruce2017}. Not only does this remain a viable explanation for the Sharov21 event, it may also be the tip of the proverbial iceberg for the archival/future detection of microlensed background AGN in the vicinity of M31 or other nearby galaxies.

Our main aim is a critical re-evaluation of the light curve data for Sharov21, and then to carry out a microlensing model fit. We also examine whether or not Sharov21 seems a normal AGN. We then go on to discuss other possible microlensing events seen through M31. In Section \ref{sec:Data} we describe the data sets used in our analysis. Section \ref{sec:Methods} details the microlensing model and MCMC model-fitting procedures. In Section \ref{sec:Results} we present our results and in Section \ref{sec:Discussion} we assess event probabilities and implications for the future.


\section{Data}
\label{sec:Data}

\begin{figure}
    \centering
    \includegraphics[scale=0.9]{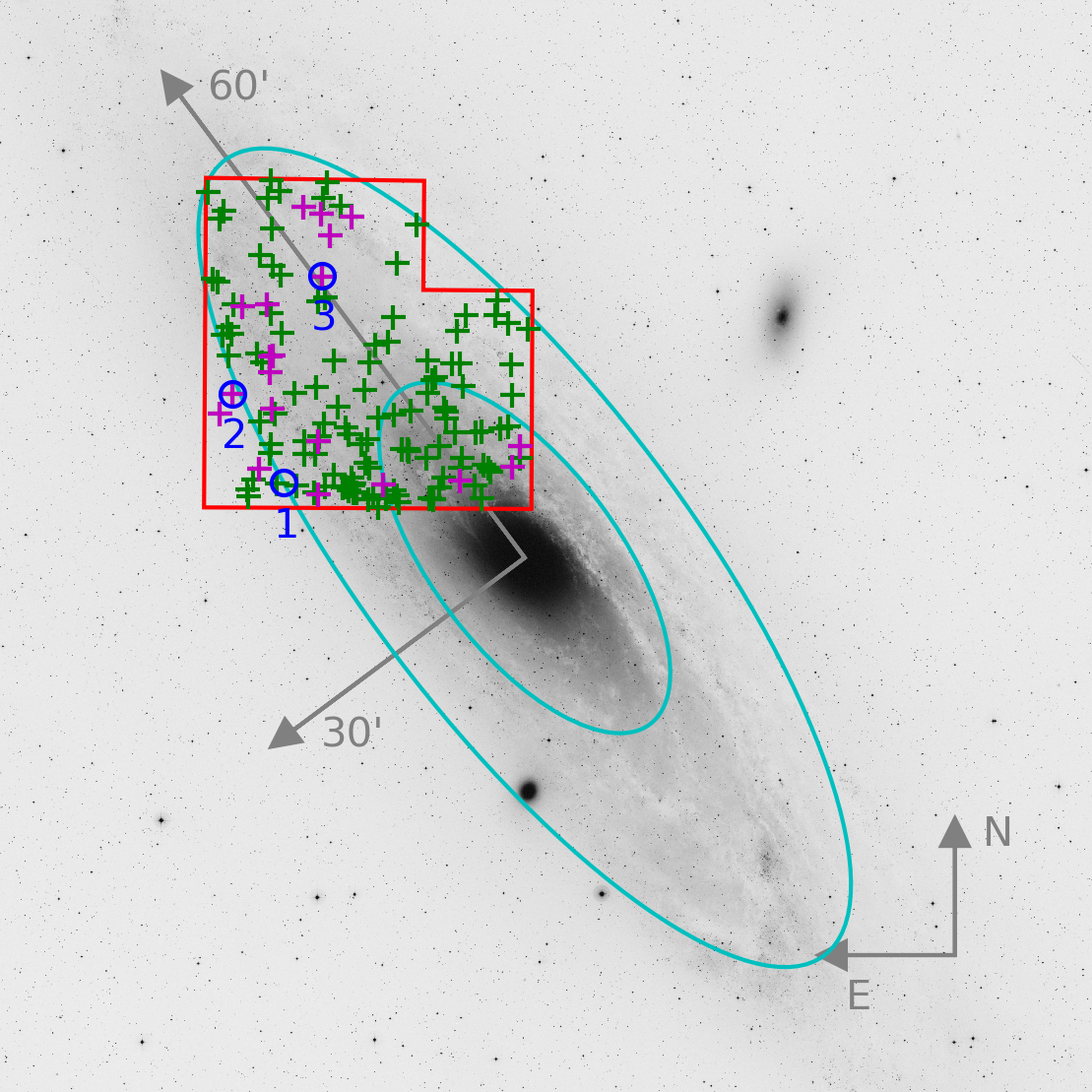}
    \caption{Wide angle view of M31. The image is a V band UK Schmidt plate (via DSS) and the major and minor axes are drawn assuming a position angle for the M31 disc of $37^\circ$ with indicative lengths of $1^\circ$ and $30\arcmin$ for the major and minor axes respectively. The figure shows the location of Sharov21 (circle labelled `1') and other objects from the \citet{Vilardell2006} survey, discussed in Section \ref{sec:Villardell}. The ellipses are isodensity contours with semi-major axes of $1300\arcsec$ and $3100\arcsec$ respectively and are discussed in Section \ref{sec:Rates}.}
    \label{fig:M31}
\end{figure}

This section provides an overview of the suite of data sets we have employed. We will describe 1) the time series data for the Sharov21 event 2) imaging data used to construct an SED for Sharov21 including HST, Spitzer and XMM-Newton observations 3) the \citet{Vilardell2006} study that we use to search for more examples of Sharov21-like events. For reference, Figure \ref{fig:M31} shows a wide angle view of M31 and the objects of particular interest to this work. The three most promising candidates for microlensing events, including Sharov21, are highlighted here along with the results of our search for additional candidate microlensing events.

\subsection{Sharov21 time series data}
\label{sec:S21timeseries}

The archival time series data for Sharov21 used in this paper are almost identical to that used in \citet{Meusinger2010}. As noted in that work, approximately 80\% of the data has been taken in the B band whilst the remaining data has been corrected to $B$. The resulting B-band light curve spans several decades and the numerous sources of data, including archival plates, are detailed therein. The sources of these data and number of epochs are listed in Table \ref{tab:S21timeseries}. The primary data set is that from \citet{Sharov1998} which comprises of data from four telescopes. We note that no formal errors are reported in that work. \citet{Meusinger2010} adopt a 0.1 mag error for this data which we carry forward though this assumption should be treated with caution. We have also opted to exclude 10 epochs flagged as being either uncertain and/or of low quality. With the addition of one further epoch from SuperCOSMOS \citep{Hambly2001}, which agrees well with the quiescent value, this gives us a total of 147 epochs. Table \ref{tab:S21timeseries} also notes the number of epochs available within 6 months of the peak of the light curve on approximately MJD 48916. It is these data that will prove to be the most important in testing the microlensing hypothesis.

\begin{table}
    \centering
    \begin{tabular}{l|l|l|l}
        Source                              & $\rm{N_{epochs}}$ & $\rm{N_{peak}}$ & Plates? \\
        \hline
        Sharov et al (1998) Table 1         & 58    & 50    & y \\
        Sharov et al (1998) 2-m Roshen      & 1     &       & y \\
        Tautenburg Schmidt photografisch    & 29    & 6     & y \\
        Tautenburg Schmidt CCD              & 14    &       &   \\
        ING Vilardell et al.                & 6     &       & y \\
        NOAO LGS                            & 2     &       &   \\
        CA 2.2-m CAFOS                      & 2     &       &   \\
        CA Schmidt                          & 14    & 4     & y \\
        CA 1.2-m                            & 5     &       & y \\
        Palomar Schmidt                     & 4     &       & y \\
        Asiago Schmidt                      & 7     & 6     & y \\
        INT (WFS)                           & 1     &       &   \\
        3.6-m CFHT                          & 1     &       &   \\
        40-cm Astrograph Sonneberg          & 1     & 1     & y \\
        60-cm Ganymede Skinakas             & 1     &       &   \\
        SuperCOSMOS                         & 1     &       & y \\
        \hline
        Total                               & 147   &   &      \\
    \end{tabular}
    \caption{Data sources for the Sharov21 time series. With the exception of the SuperCOSMOS data, the source IDs are noted as per the data supplied to us and used in \citet{Meusinger2010}.}
    \label{tab:S21timeseries}
\end{table}

\begin{figure}
	\includegraphics[width=\columnwidth]{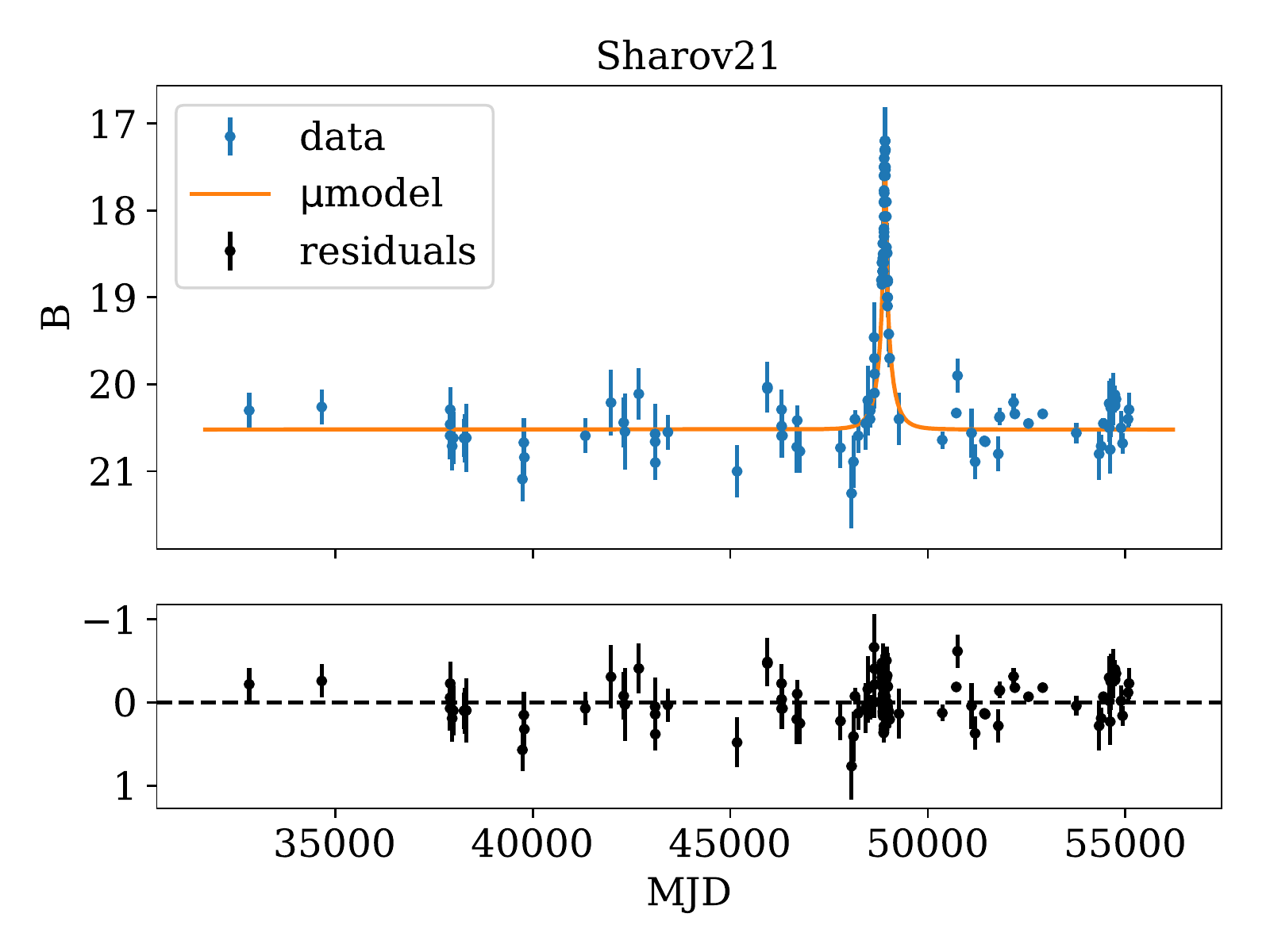}
    \caption{Sharov21 full light curve including our microlensing model (discussed in Section \ref{sec:S21microResults}). Residuals plotted below. Error bars reflect the original errors on the photometry.}
    \label{fig:S21_LC}
\end{figure}

\begin{figure}
	\includegraphics[width=\columnwidth]{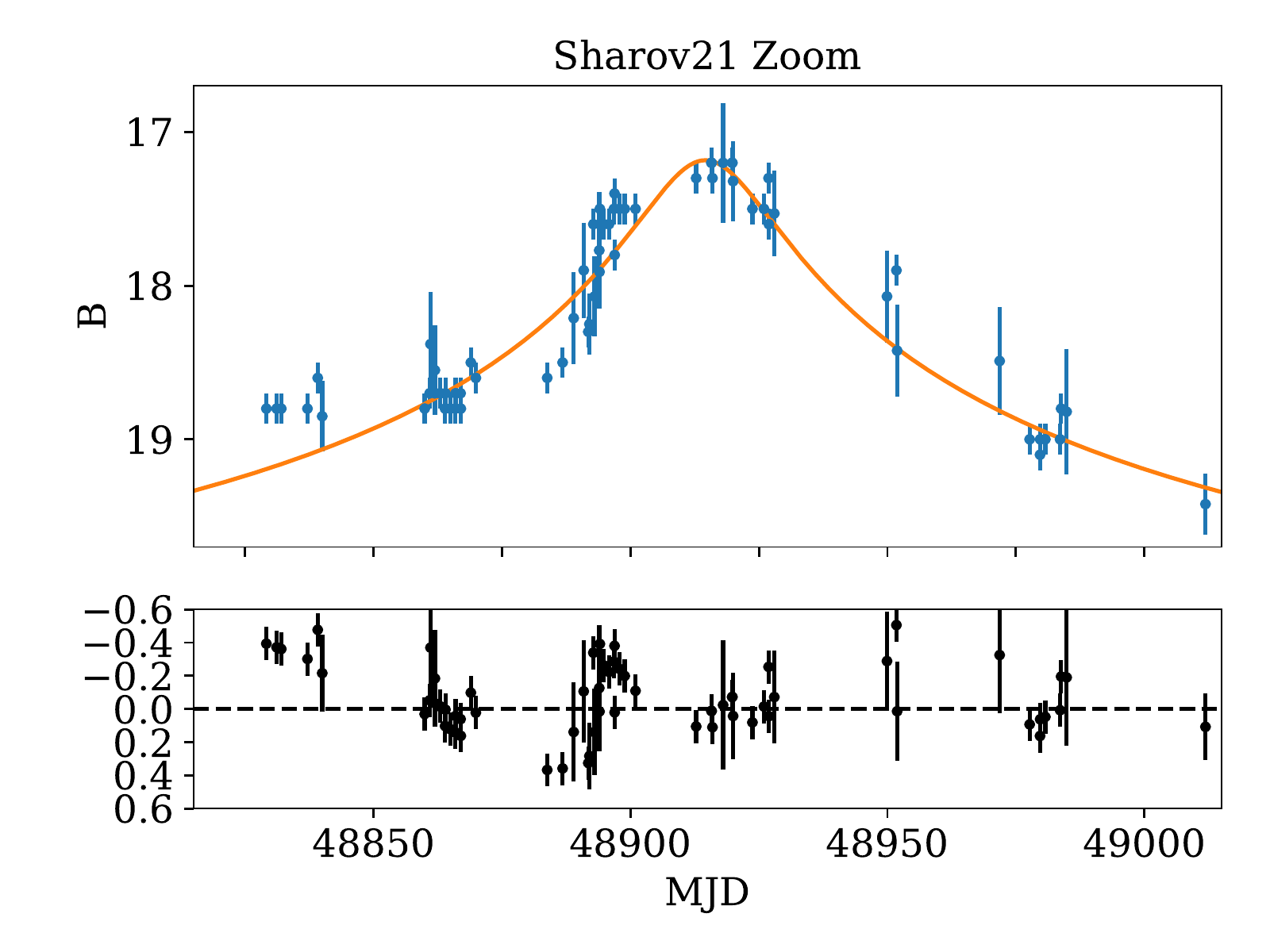}
    \caption{Sharov21 zoomed light curve including our microlensing model (discussed in Section \ref{sec:T2microResults}). Residuals plotted below. Error bars reflect the original errors on the photometry.}
    \label{fig:S21_LCzoom}
\end{figure}

This impressive decades-long lightcuvre, shown in Figure \ref{fig:S21_LC}, allows us to confidently state that Sharov21 shows no sign of any notable outburst with the exception of the main event in 1992. As reported in \citet{Meusinger2010}, there are two other things to note with regards the light curve data. The first is that there is tentative evidence for a more rapid rise to the peak, as evidenced by a shoulder in the data, which can also be seen in our zoomed view of the light curve (Figure \ref{fig:S21_LCzoom}). The second is that there are reported colour changes \citep[see][Fig. 3]{Meusinger2010}. Both of these observations, if accurate, call into question the microlensing hypothesis. This is because, particularly in the case of a point-source, microlensing events are expected to be achromatic in nature and display a symmetric light curve.

The data showing evidence for the more rapid rise to peak comes from the Sharov plates, primarily those taken at their Crimean station. Given our incomplete understanding of the systematics inherent in this photographic plate data, the assumed 0.1 mag errors may be overly optimistic. The third-party, non-Sharov observational data about the peak (data points in Figure \ref{fig:S21_LCzoom} with larger errors), appear to be in good agreement with the model shown and this reinforces the notion that the shoulder may simply be an artefact in the data rather than a real change. With reference to the previously reported colour indices, it is only the B-R data that show colour information for epochs with B < 20. Note also that, though there is an indication of a trend, the B-R data is still consistent with being flat, i.e.: it is not sufficient to rule out the achromatic case.

With these caveats in mind, we retain the assumed 0.1 mag errors on the Sharov data for our microlensing analysis. We cannot rule out the presence of the shoulder or colour changes but neither do we believe that the data quality is sufficient to rule out the null hypothesis. We continue with the assumption that the simple point-lens, point-source microlensing model remains valid in the case of Sharov21.

\subsection{Optical imaging data}
The Panchromatic Hubble Andromeda Treasury (PHAT) survey \citep{Dalcanton2012} with the Hubble Space Telescope (HST) provides high-resolution imaging in the UV (WFC3/UVIS F275W, F336W), optical (ACS/WFC F435W, F814W) and near-infrared (WFC3/IR F110W, F160W). We construct a catalogue of all objects in the PHAT brick 8, field 14, imaging using the software {\tt SExtractor} \citep{Bertin1996} in dual-image mode with the F814W imaging serving as the detection image.

All of the photometry for the HST imaging was measured with $0.4^{\prime\prime}$-diameter apertures, and corrected to total assuming a point-source correction, as estimated from the flux versus radius curve-of-growth. Although the aperture correction to total is more sizeable for the F160W filter -- a factor $\simeq 1.7$ compared to the more modest factor $\simeq 1.2$ for the UV and optical filters -- this is necessary given the severe source crowding within the field. Moreover, this particular object can also be reasonably approximated to be a point-source.

Photometric errors for each HST filter were calculated using a local depth analysis (e.g. \citet{McLeod2016}). A grid of non-overlapping $0.4^{\prime\prime}$ apertures were placed spanning the entire field of view and, using {\sc SExtractor}, a segmentation map was generated in order to mask out any apertures containing significant flux from sources. Statistical analysis was performed on a smaller grid of $\simeq 150$ ``blank-sky'' apertures local to our target object, and the Median Absolute Deviation (MAD) estimator was used in order to derive the local $\sigma$ photometric error.

To supplement our HST photometry, we also included the photometry in {\it g, r, i, z} and {\it y} from the DR1 release of the Panoramic Survey Telescope and Rapid Response System (Pan-STARRS), \citep{Chambers2016,Magnier2016a,Magnier2016b,Magnier2016c,Waters2016,Flewelling2016}. Here we use the PSF magnitudes.

\begin{figure}
	\includegraphics[width=\columnwidth]{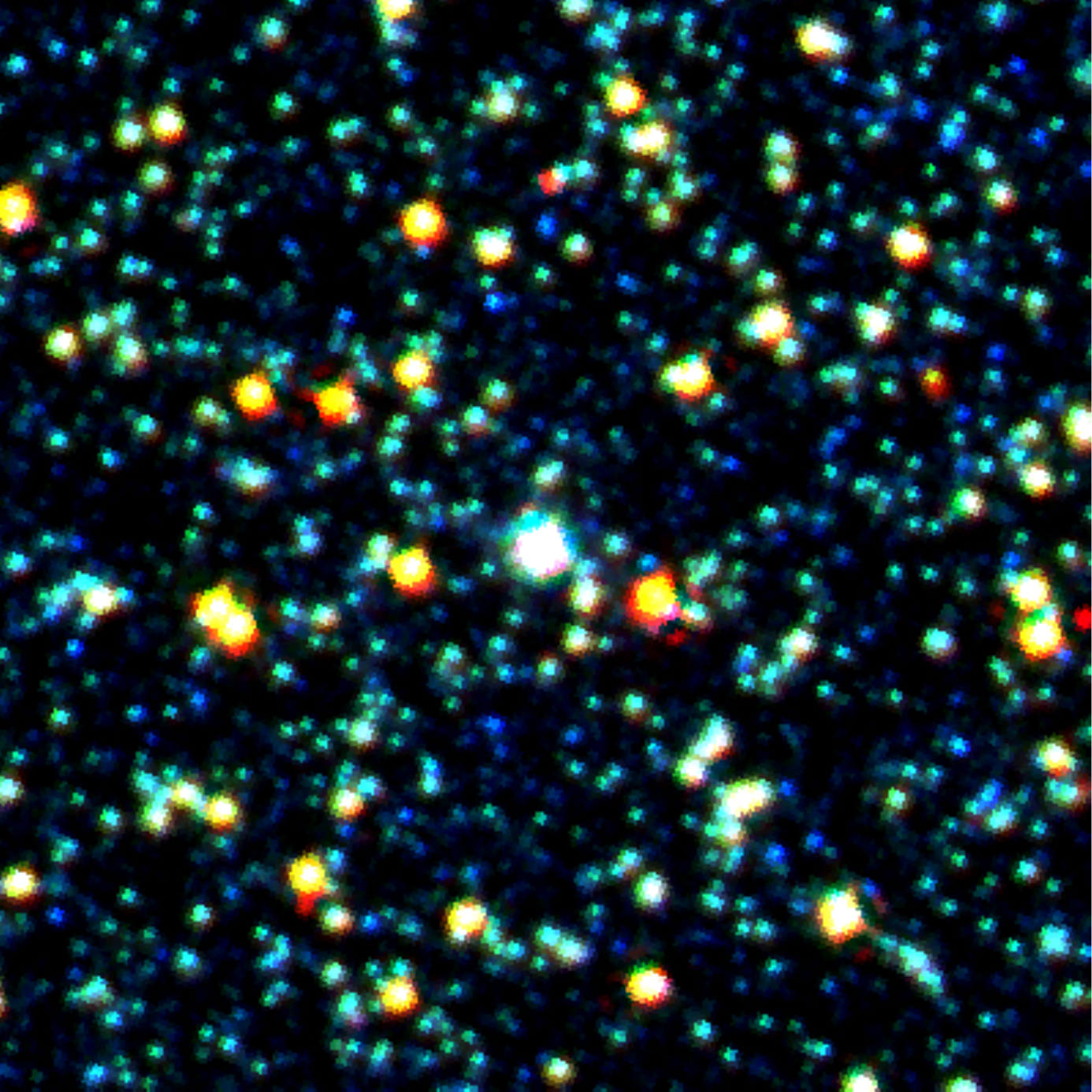}
    \caption{Colour composite of a $10^{\prime\prime}$x$10^{\prime\prime}$ region centred on Sharov21 using our stacks of the PHAT survey data. Blue channel: F475W filter; green: F814W; red: F110W.}
    \label{fig:S21_cutout}
\end{figure}

Figure \ref{fig:S21_cutout} shows a composite cutout of a $10^{\prime\prime}$x$10^{\prime\prime}$ region surrounding Sharov21, approximately $26^\prime$ away from centre of M31. This false colour image makes use of our PHAT stacked data in three filters: F475W, F814W and F110W and provides an excellent combination of resolution and sensitivity. Sharov21 is well described by a point source in these images. At longer wavelengths blending becomes a serious concern which makes an accurate background subtraction difficult, as will be described in Section \ref{sec:SED}.

\subsection{Mid-infrared data}
Sharov21 has also been imaged in several Spitzer programs\footnote{The program IDs are 3400, 3126, 61001} and we include mid-infrared photometry in our analysis. Imaging in the mid-infrared is from the Spitzer Infrared Array Camera (IRAC), downloaded from the IRSA Spitzer Heritage Archive. This includes imaging in channel 1 ($3.6\,{\rm \mu m}$), channel 2 ($4.5\,{\rm \mu m}$), channel 3 ($5.8\,{\rm \mu m}$) and channel 4 ($8.0\,{\rm \mu m}$). One drawback of Spitzer imaging is the blending of sources due to the broad PSF FWHM, and so one requires deconfusion techniques to extract reliable photometry. To this end, we used the deconfusion software {\tt TPHOT} \citep{Merlin2015}. This uses the spatial and surface brightness information based on the high-resolution imaging (HST F814W in this instance) as a prior. Cutouts of all objects in an input catalogue are then convolved with a kernel which produces object model templates in the low-resolution IRAC image. The fluxes of all these templates are then fitted simultaneously. In order to be consistent with the aperture-corrected photometry, we opt to subtract all objects except the target Sharov21, and then perform aperture photometry on the cleaned-up image to avoid contamination. For channel 1 and channel 2, we employ $2.8^{\prime\prime}$-diameter apertures, and for channel 3 and channel 4, we use $5^{\prime\prime}$-diameter apertures, before correcting to the total flux assuming a point-source and the estimated flux curve-of-growth.

\subsection{X-ray and radio data}
The XMM-Newton data have been taken directly from the XMM-Newton Science Archive. Of four XMM observations, two had a reasonable number of counts and the catalogue fluxes for these have been averaged across each bandpass. Assumed upper limits for GALEX and two M31 radio surveys are taken directly from \citet{Meusinger2010} \citep{Braun1990,Gelfand2004}.

\subsection{Vilardell survey time series data}
\label{sec:Villardell}

The Sharov21 event may not be the only one of its kind so, in order to begin a search for other candidate events, we require suitable long-term monitoring data in the immediate vicinity of M31. The cadence of this data need not be high as we are primarily concerned with timescales on the order of months or more.

A very useful archival source of time series data is that from \citet{Vilardell2006}. This four year photographic survey of the Northeastern quadrant of M31 of comprises approximately five month-long observing epochs in both B and V, each separated by one year. Our Sharov21 lightcurve includes binned data from this survey. Though the survey was designed to search for variable stars, in particular eclipsing binaries, it nevertheless allows us to search for additional long term transient events which may be similar in nature to Sharov21.

Out of their 236,238 sources, 3,964 have been identified as a variable star over an approximate $34\arcmin\times34\arcmin$ field of view. Of these, 853 have been classified as an eclipsing binary or Cepheid. The remaining 3,111 object catalogue contains other variables with periods outwith the 1-100 day analysis range and also includes non-periodic sources. It is this `variable star' catalogue that we make use of to perform a search for other candidate AGN microlensing events (see Section \ref{sec:search}).

\section{Methodolgy}
\label{sec:Methods}
\subsection{Microlensing analysis}

With the light curve noted in the previous section, we take a similar approach to modelling the lensing event as \citet{Bruce2017}. Our first assumption is that the event can be well described by a simple microlensing model of a point-mass object passing in front of a background point-source. The light curve for which has an analytic solution involving the following parameters: source redshift ($z_{\rm s}$); lens redshift/mass/transverse velocity ($z_{\rm d}, M_{\rm l}, v_\perp$); impact parameter ($y_0$); mid-point epoch ($t_0$); source/background flux ($F_{\rm s}, F_{\rm b}$). With the source redshift in this case constrained to be $z_{\rm s}=2.109$ from spectroscopy this leaves 7 free parameters, also noted in Table \ref{tab:microParams}. Cosmological calculations in this paper make use of \texttt{Planck13} values: \citet{Planck2014}; $H_0=67.8, \Omega_\Lambda=0.693$.

\begin{table}
	\centering
	\caption{Free parameters in the simple microlensing model.}
	\label{tab:microParams}
	\begin{tabular}{cl}
		\hline
		Parameter & Description \\
		\hline
		$M_\textrm{l}$ & Lens mass \\
		$v_\perp$ & Transverse velocity \\
		$t_0$ & Mid-point epoch \\
		$z_\textrm{d}$ & Lens redshift \\
		$y_0$ & Impact parameter \\
		$F_\textrm{s}$ & Source flux (pre-lensing) \\
		$F_\textrm{b}$ & Background flux (unlensed) \\
		\hline
	\end{tabular}
\end{table}

In order to efficiently explore this parameter space we employ the software package {\tt emcee} \citep{Foreman-Mackey2013} to perform an MCMC analysis. We also include some relatively simple assumptions in the choice of priors. This includes a log-normal prior on the lens mass ($\mu=0, \sigma=1$, median value 1\,M$_\odot$) to ensure we are in the stellar-mass regime and a gaussian prior on the transverse velocity, centred on 400 km/s with a sigma of 200 km/s. A further consideration was to place a constraint of $\pm$60 days about the peak to minimise the amount of time spent in low likelihood regions.

We know that any AGN will also display some level of intrinsic variability regardless of any extrinsic cause. To allow for this in the MCMC analysis, we increased the errors on the light curve as $\sigma_{\rm MCMC}^2=\sigma_{\rm phot}^2 + \sigma_{\rm AGN}^2$, where we initially set $\sigma_{\rm AGN}$=0.1, a conservative value for typical AGN fluctuations.

Initial testing with a starting guess for the lens redshift of $z_{\rm d}=0.2$ failed to converge in the time allotted. With the possibility that the true lens position may lie at the much smaller ``redshift'' corresponding to M31 (on the order $z\sim 10^{-4}$) the decision was made to perform a search in log($z_{\rm d}$) instead. With the starting guess kept the same, this second attempt converged successfully. The results of the analysis are shown in Section \ref{sec:S21microResults} with parameter constraints taken using the one-sigma percentiles from the MCMC trace and the `best-fit' model in this case corresponding to the peak in the posterior distribution.

\subsection{Search for other microlensing candidates}
\label{sec:search}

In order to search for additional microlensing candidates which may be similar to Sharov21, we perform a search through the published \citet{Vilardell2006} data tables. In particular, their sources already identified as variable but not confirmed as either an eclipsing binary or Cepheid, a total of 3,111 objects. The data comprises groupings of approximately month-long observing periods, performed once per year, over a total four years. As we are concerned primarily with locating candidate events with $\sim$year-long timescales, the median magnitude across each of the five principal observing epochs was determined after excluding data with uncertainty > 0.5 mag. We then exclude objects classified as variable stars and perform a search for changes of > 0.75 mag in either B or V across any of the five epochs. To exclude sources displaying rapid variability, an additional constraint was that the standard deviation over any one of the five clusters of observations was not more than 0.2 mag. After imposing these constraints, we identified 139 objects of interest. Each lightcurve was then visually inspected for smooth variations over the period of observations, as would be expected in a point-source, point-lens microlensing event. The assumption that the event need be achromatic was relaxed during this process. This left us with 20 candidate events. Of these 20, two objects were identified as likely background AGN due to the existence of corroborating data. The candidates are plotted in Figure \ref{fig:M31} with position information noted in Table \ref{tab:TargetIDs}.

T2 has spectroscopic data which confirms the presence of a background AGN at $z=0.215$ \citep{Dorn-Wallenstein2017}. It has been identified as a possible binary AGN though the binary nature of this AGN is in doubt \citep{Barth2018}. Nevertheless, the lightcurve for this object underwent a smooth, significant change of $\sim 0.75$\,mag and has not been seen to exhibit this type of behavior in the observations since. As target T2 has a spectroscopic redshift, we perform a microlensing analysis of the event as per the procedures outlined for Sharov21. T3 has no available spectroscopy but is detected in the XMM-Newton Science Archive (as are all our targets) which makes it likely that this is another background AGN. In this case the lightcurve is seen to undergo a smooth, significant change of $\sim 1.1$\,mag. Without a spectroscopic redshift, we do not include T3 in our MCMC microlensing analysis.

\begin{table}Target IDs/positions
    \centering
    \begin{tabular}{c|c|c|c|c|c}
    LongID & ShortID & RA & Dec \\
    Sharov21 & S21 & 00:44:57.94 & +41:23:43.72 \\ 
    M31V\_J00452730 & T2 & 00:45:27.31 & +41:32:54.06 \\ 
    M31V\_J00443792 & T3 & 00:44:37.95 & +41:45:14.10 \\
    M31 centre & M31* & 00:42:44.35 & +41:16:08.63 \\
    \end{tabular}
    \caption{Target coordinates for the three microlensing candidates and our assumed position for the centre of M31. T3 coordinates are from the XMM-Newton Science Archive detection, within 0.3 arcsec of the \citet{Vilardell2006} position. Sharov21 and T2 are GAIA coordinates.}
    \label{tab:TargetIDs}
\end{table}

\section{Results}
\label{sec:Results}

\subsection{Sharov21 SED}
\label{sec:SED}

\begin{figure}
	\includegraphics[width=\columnwidth]{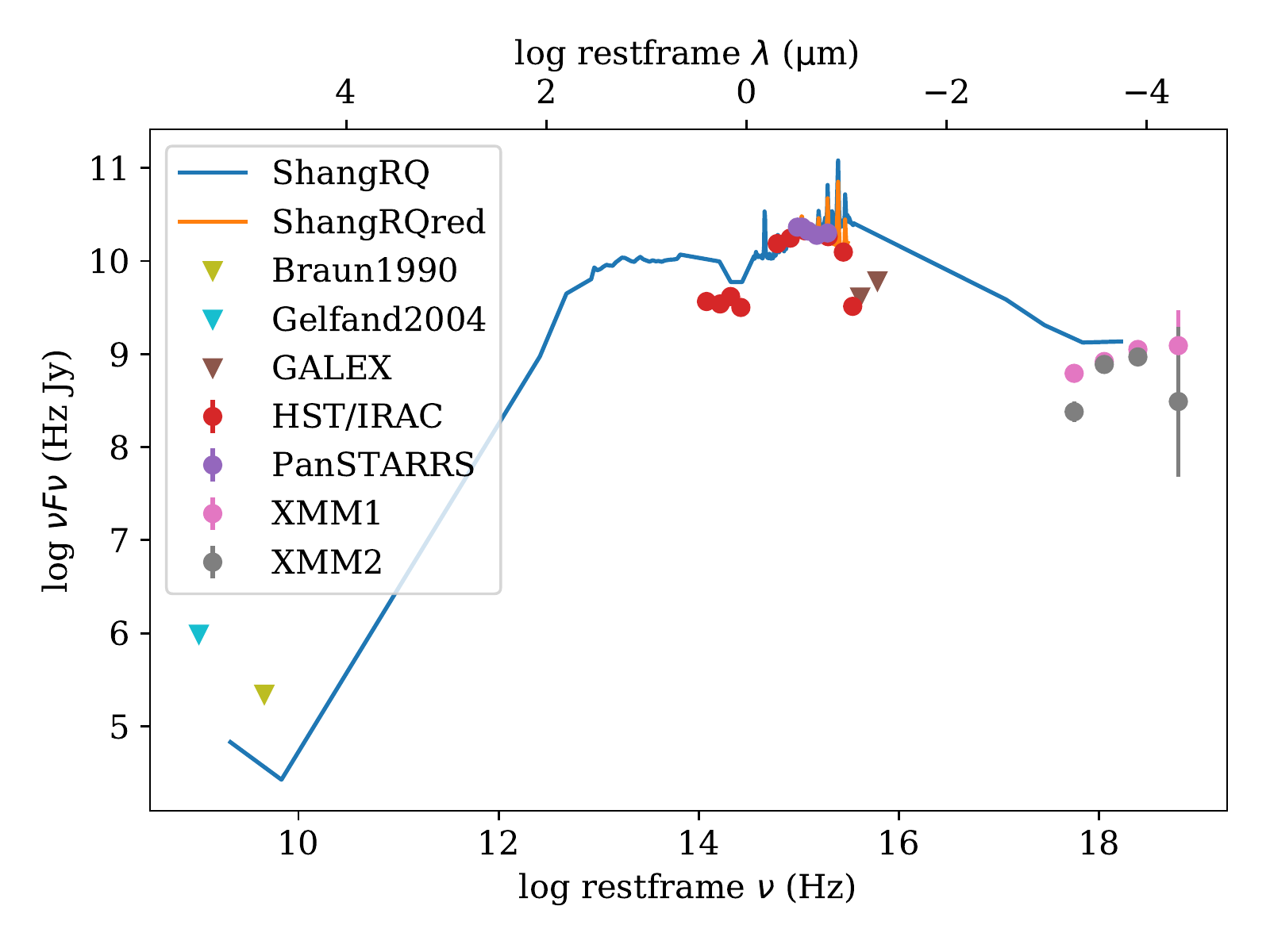}
    \caption{Full SED for Sharov21. The data points have not been corrected for Milky Way reddening. The triangles represent upper limits.}
    \label{fig:S21_SEDfull}
\end{figure}

\begin{figure}
	\includegraphics[width=\columnwidth]{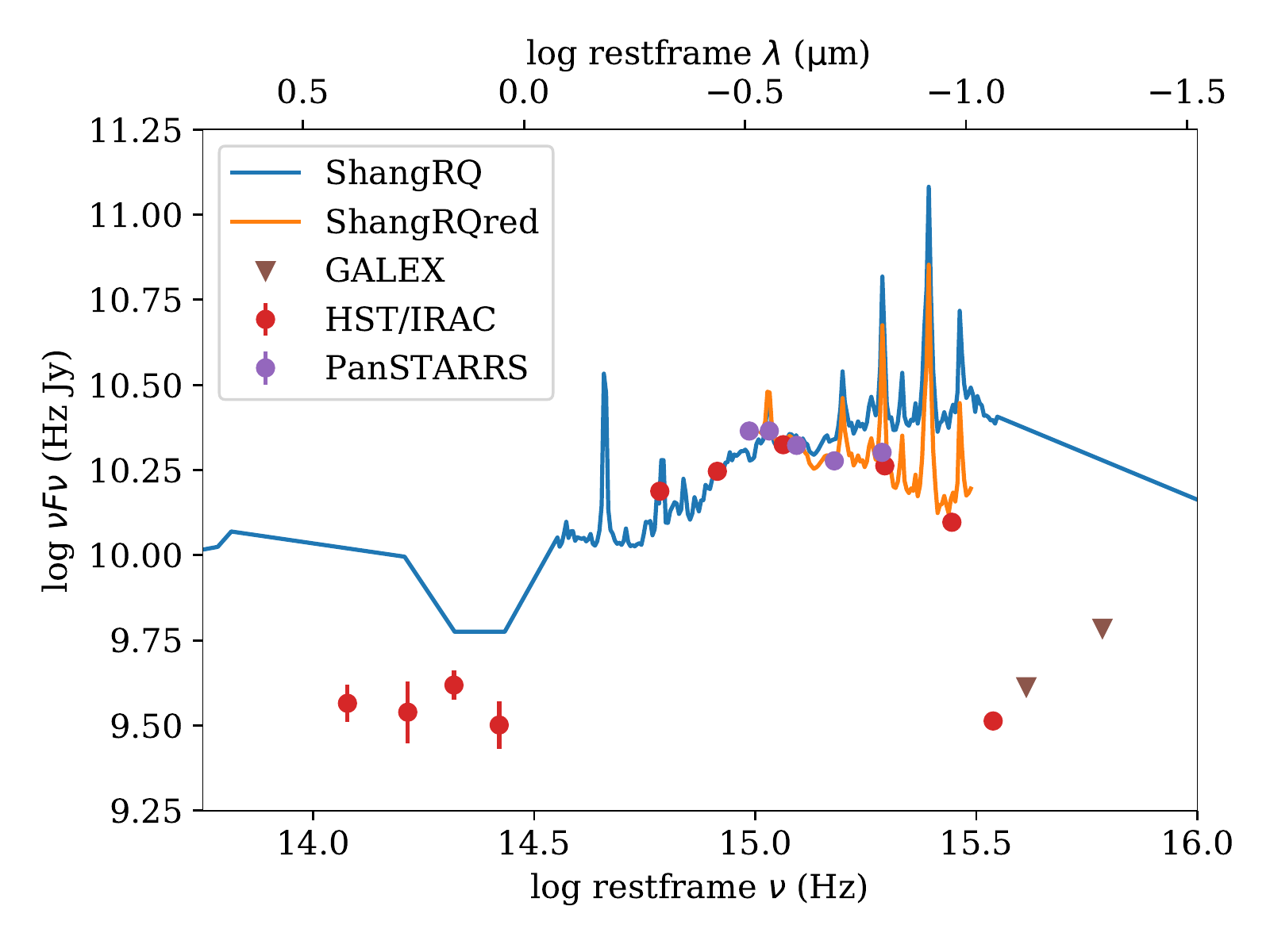}
    \caption{Zoom in on the IR--UV range of the SED for Sharov21, the data points have not been corrected for Milky Way reddening though the orange line reflects the \citet{Shang2011} template, reddened assuming $B-V=0.2$}
    \label{fig:S21_SEDzoom}
\end{figure}

The full SED is shown in Fig. \ref{fig:S21_SEDfull}. The first point to note is that at there is nothing immediately unusual about this AGN. It is clear from the non-detections in the radio that this object can be considered radio-quiet and, for comparison, we have included the radio-quiet SED template from \citet{Shang2011} which is in broad agreement. At present, we have made no attempt to correct for Milky Way reddening (approximately $B-V=0.2$) which explains the drop-off seen in the UV. In Fig. \ref{fig:S21_SEDzoom} we show a zoomed in view on the IR-UV region which includes a reddened SED template that provides a good match to the data. One potential issue is that the IRAC mid-IR fluxes seem to be underestimated by $\sim 40\%$, most likely to do with the uncertainties inherent in working in a crowded field such as this. We believe our estimates are more accurate than the published Spitzer Enhanced Imaging Product (SEIP) source list values for this object. This is particularly true for IRAC channel one which appears anomalously high at the expected rest-frame $1\mu{\rm m}$ minimum, perhaps a problem concerning contamination from one or more neighbors. Indeed, the SEIP source list entry is flagged as having a bad background match.

\subsection{Microlensing analysis results for Sharov21}
\label{sec:S21microResults}

\begin{table}
	\centering
	\label{tab:MCMC_S21}
	\begin{tabular}{clc}
    	\hline
		Sharov21 & & $z_{\rm agn}=2.109$ \\
		\hline
		parameter & value & unit \\
		\hline
		${\rm log_{10}}(z_{\rm d})$	& $-3.82\,^{+0.63}_{-0.67}$ & \\
		$M_{\rm l}$					& $1.26\,^{+2.39}_{-0.79}$ & M$_\odot$ \\
		$v_\perp$					& $420\,^{+205}_{-192}$ & km\,s$^{-1}$ \\
		$y_0$						& $0.0343\,^{+0.0077}_{-0.0075}$ & $\theta_E$\\
		$t_0$						& $48915.7\,^{+0.9}_{-0.9}$ & MJD \\
		$F_{\rm s}$					& $2.52\,^{+0.55}_{-0.54}$ & $\times10^{-17}$erg\,s$^{-1}$cm$^{-2}$\AA$^{-1}$ \\
		$F_{\rm b}$					& $9.20\,^{5.14}_{-5.30}$ & $\times10^{-18}$erg\,s$^{-1}$cm$^{-2}$\AA$^{-1}$ \\
		$r_{\rm E}$					& $1260\,^{+2780}_{-860}$ & light-days \\
        \hline
	\end{tabular}
	\caption{Results of the MCMC microlensing analysis for Sharov21. The full corner plot can be found in Fig. \ref{fig:S21_corner}.}
\end{table}

We first report on our analysis of the Sharov21 event. The results from the MCMC analysis are displayed in Table \ref{tab:MCMC_S21} and the corresponding corner plot is displayed in Fig. \ref{fig:S21_corner}. In general, the parameters are well constrained. Of particular interest is the range of allowable values for the projected lens redshift. These redshifts correspond to a physical distance in the range 0.14--2.84\,Mpc with the most probable value being 0.67\,Mpc. This is very close to the true distance of M31, $\simeq0.78$\,Mpc.

In contrast to the suspected microlensing events reported in \citet{Lawrence2016a} and \citet{Bruce2017} the timescale for this event is quicker, with the Einstein timescale $t_{\rm E}\approx 1$\,year. This is a natural consequence of the low lens redshift but it does mean that events of this kind are in a more favourable regime for a coordinated observation campaign if detected on the rise. The peak amplification of the event is a factor of 30 above the base AGN level of ${\rm B}\sim20.5$\,mag. A further consequence of the low lens redshift is that the Einstein radius when projected at the distance of the source ($r_{\rm E}$) is on the order of light years as opposed to light days. With this lens footprint, our assumption that the AGN can be regarded as a point source would appear to be secure as any radial temperature profile in the accretion disc would be expected to go unresolved. Chromatic effects would be expected to creep in if the source is larger than approximately 10\% of the Einstein radius. This may also mean that the inner regions of the BLR may have undergone significant amplification during this event.

How well does the microlensing model explain the bulk changes seen in this object? The full light curve is displayed in Fig. \ref{fig:S21_LC} and a zoomed-in view of the main event is displayed in Fig. \ref{fig:S21_LCzoom}. In both cases we overplot the model which corresponds to the peak of the posterior distribution. This model, using our MCMC errors defined above, produces a reduced chi squared fit of $\chi_\nu=1.48$ and broadly speaking performs very well. A potential issue can be seen in the residuals to the zoomed-in view where the data, particularly around the ${\rm MJD}\sim48880$ mark, shows some structure. As discussed in Section \ref{sec:S21timeseries} we believe that this shoulder in the data may be unreliable. Tests with simulated damped random walk (DRW) models (e.g.: \citet{MacLeod2012}), using typical parameters for radio-quiet AGN, show that changes of this nature occur very rarely, if ever, when using the same cadence/sampling as the Sharov data. However, a modest increase to our intrinsic variability parameter, $\sigma_{\rm AGN}=0.15$, is sufficient to bring the reduced chi squared value of the fit to unity.

\subsection{Microlensing analysis results for candidate T2}
\label{sec:T2microResults}

\begin{figure}
	\includegraphics[width=\columnwidth]{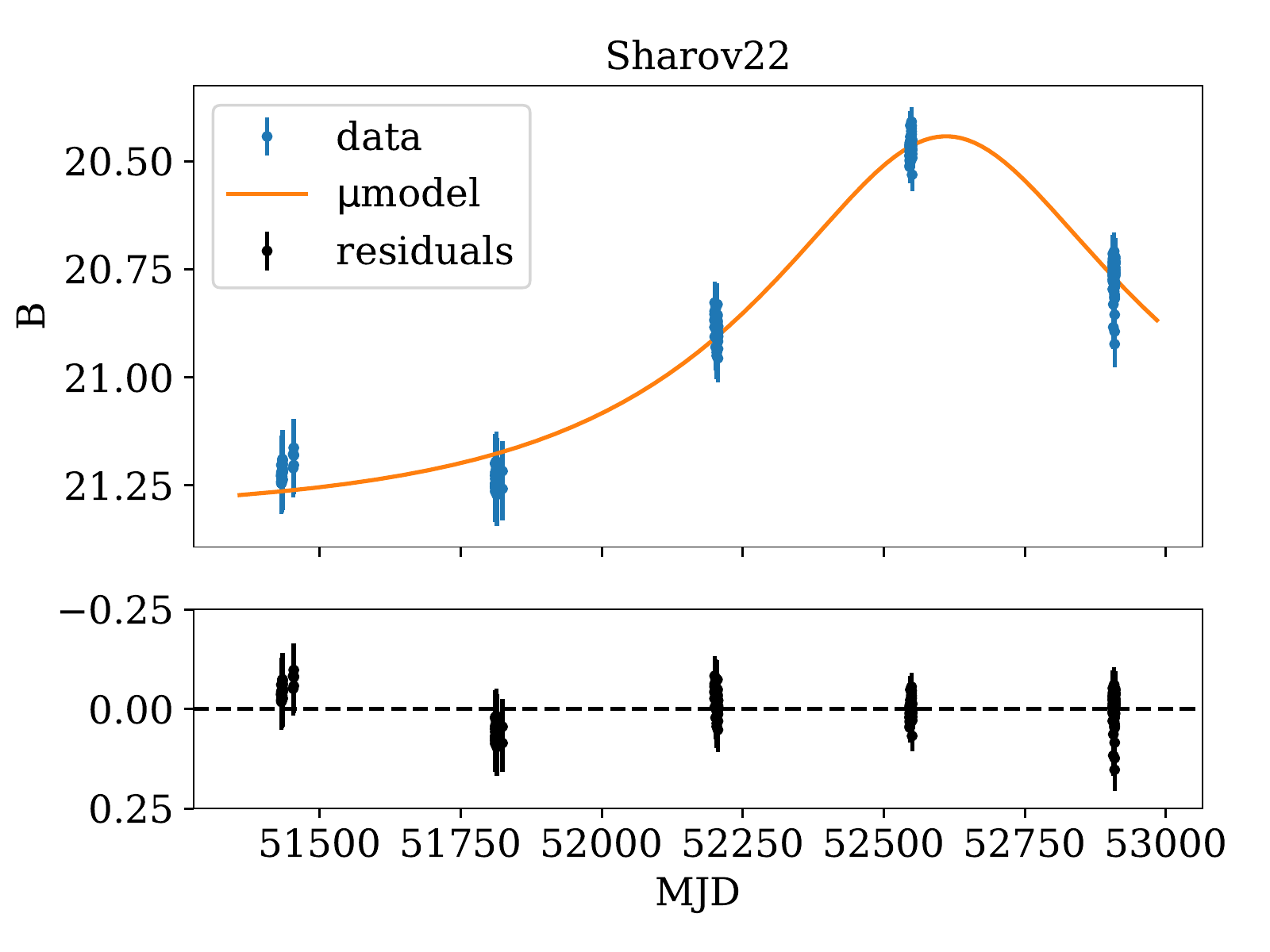}
    \caption{T2 light curve including our microlensing model. Residuals plotted below. Error bars reflect the original errors on the photometry.}
    \label{fig:T2_LC}
\end{figure}

In addition to the Sharov21 analysis, we have also performed a microlensing analysis of the candidate T2 event. The results are reported in Table \ref{tab:MCMC_T2} and the corresponding fit is shown in Figure \ref{fig:T2_LC}. It is immediately apparent that the sampling of the data is sub-optimal but the fit to the data performs very well with a reduced chi squared fit of $\chi_\nu=0.15$. The redshift values correspond to a physical distance in the range 0.63--11.0\,Mpc with the most probable value being 2.63\,Mpc. This is greater than the true distance to M31 though is still consistent with the lens residing in M31 within the one sigma constraints noted. In contrast to the Sharov21 event, the peak amplification is much lower, a factor of two above the background level in the B-band.

\begin{table}
	\centering
	\label{tab:MCMC_T2}
	\begin{tabular}{clc}
    	\hline
		T2 & & $z_{\rm agn}=0.215$ \\
		\hline
		parameter & value & unit \\
		\hline
		${\rm log_{10}}(z_{\rm d})$	& $-3.23\,^{+0.62}_{-0.62}$ & \\
		$M_{\rm l}$					& $1.12\,^{+1.68}_{-0.67}$ & M$_\odot$ \\
		$v_\perp$					& $420\,^{+203}_{-187}$ & km\,s$^{-1}$ \\
		$y_0$						& $0.424\,^{+0.045}_{-0.097}$ & $\theta_E$\\
		$t_0$						& $52612\,^{+8}_{-8}$ & MJD \\
		$F_{\rm s}$					& $1.37\,^{+0.21}_{-0.40}$ & $\times10^{-17}$erg\,s$^{-1}$cm$^{-2}$\AA$^{-1}$ \\
		$F_{\rm b}$			    	& $<6.5$ & $\times10^{-18}$erg\,s$^{-1}$cm$^{-2}$\AA$^{-1}$\\
		$r_{\rm E}$					& $249\,^{+484}_{-167}$ & light-days \\
        \hline
	\end{tabular}
    \caption{Results of the MCMC microlensing analysis for T2. The full corner plot can be found in Fig. \ref{fig:T2_corner}.}
\end{table}

This object displays differential evolution in the lightcurve, with a change in the V-band of $\sim0.25$ mag. This differential change is more significant than the tentative colour change noted for Sharov21 however this does not immediately rule out microlensing as the cause of the event. The extended nature of the T2 host galaxy is clear in the PHAT imaging and indicates that there may be a significant contribution to the background flux in the V-band. In the B-band, the data is not sufficient to reliably constrain the background contribution and is consistent only with an upper limit which suggests that the AGN is the dominant contribution in this filter. Another possibility for the achromatic lightcurve is that the simple point-source approximation is not appropriate in this case. The constraints on the Einstein radius of the lens when projected in the source plane are smaller than that seen in Sharov21 and may be an indication that we need to allow for an extended source in order to reproduce the observed colour changes as a consequence of the accretion disc being partially resolved by the lens.



\section{Discussion}
\label{sec:Discussion}

\subsection{Evidence for microlensing}

\begin{figure}
    \centering
    \includegraphics[width=\columnwidth]{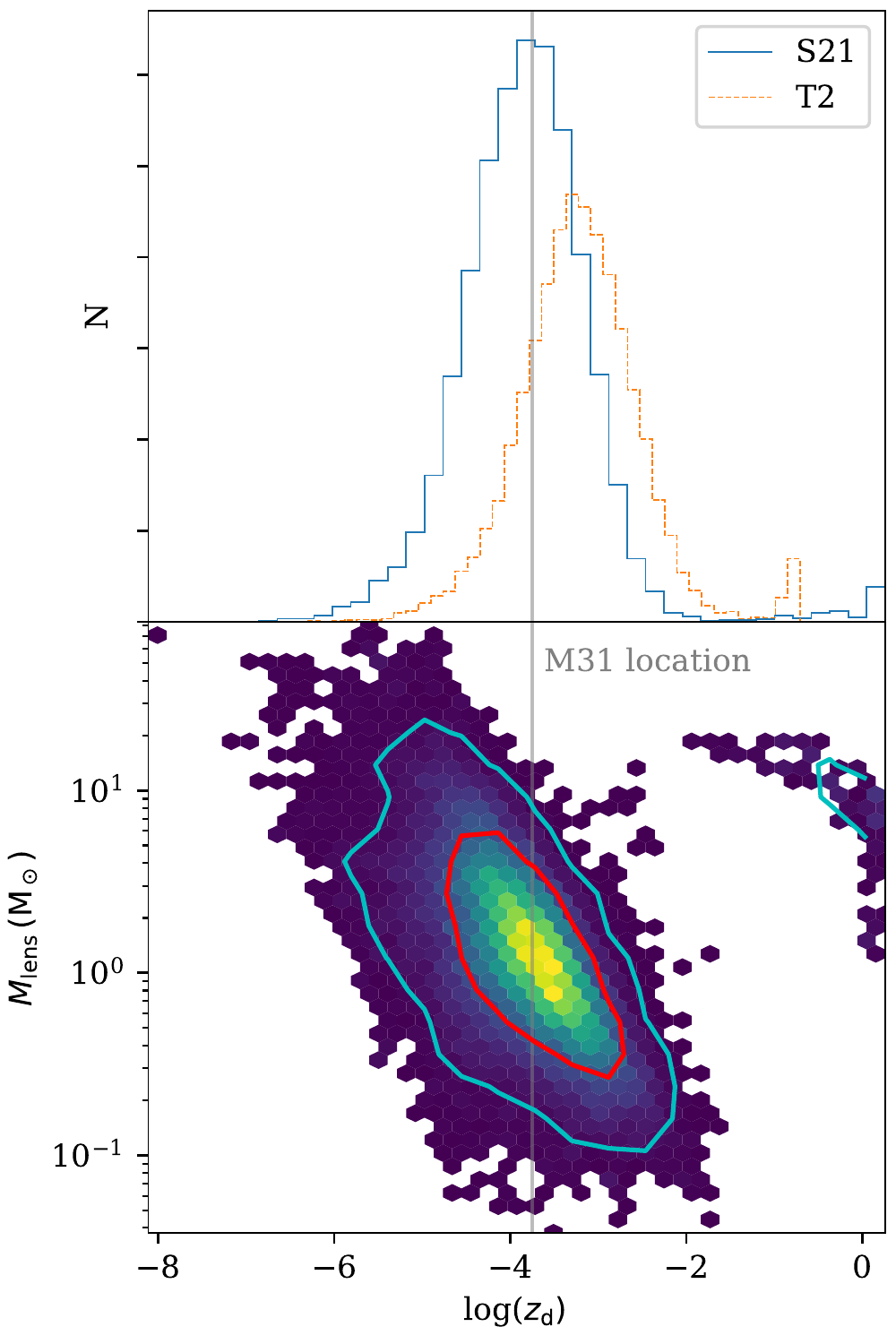}
    \caption{Top panel: 1D posterior redshift distributions for targets S21 (solid line) and T2 (dotted line) from our microlensing analysis. Bottom panel: 2D posterior in the log($z_{\rm{d}}$) $M_{\rm{lens}}$ plane for Sharov21. The solid vertical line indicated the effective ``redshift'' for M31, consistent with a distance of 785 kpc.}
    \label{fig:cropcorner}
\end{figure}

We first turn our attention to Sharov21 and the case for this being a high amplitude microlensing event. Our MCMC analysis lends strong support to the microlensing hypothesis in that it successfully predicts that the most probable lens location is at the distance to M31 with only a small number of initial assumptions. Figure \ref{fig:cropcorner} shows the posterior distributions in log($z_{\rm{d}}$) and 2D distribution in the log($z_{\rm{d}}$) $M_{\rm{lens}}$ plane. Also shown is the effective redshift that corresponds to M31 which is in excellent agreement with the data. For comparison top panel also shows the posterior distribution obtained for T2. In this case the peak of the distribution is at a greater redshift but the location of M31 is still enclosed within the one sigma bounds about this peak.

In addition to a satisfactory distance estimate, our lensing model also provides evidence that this microlensing event should be achromatic given the constraint on the projected Einstein radius at the source. As has been previously mentioned, there is reported evidence of a colour change and the presence of a shoulder in the data that is not well described by the model. However, we do not believe that the lightcurve data about the peak is of sufficient quality to falsify the simple achromatic microlensing model. As discussed in Section \ref{sec:S21microResults}, a modest increase to the error bars in our lightcurve provides for a satisfactory fit to the data. We note that it is only the Sharov observations which appear to be in tension -- all of the additional third-party data about the peak is in good agreement with the model. A better understanding of any potential systematics inherent in the Sharov plates and assumed errors would be required before discarding the simple lensing model in favour of, for example, a binary lens configuration.

The T2 event exhibits a satisfactory microlensing fit to the B-band data but the simple model is lacking in that it cannot reproduce the differential colour changes seen in the data, in particular with respect to the smaller changes in the V-band. This may be due to the presence of a significant host galaxy component resulting in a dilution of the microlensing signal. This is reinforced by the spectroscopic component fits in \citet{Dorn-Wallenstein2017} which show that in the V-band the quasar component is on the order 20\% of the total contribution in that band. We may also need to take into account the possibility that the accretion disc in this case may be being partially resolved by the lens, leading to the differential changes seen. Figure \ref{fig:cropcorner} shows that the expected lens distance does not align well with the position of M31 though this location is still reasonable given the spread in the data.

\subsection{Event Rates}
\label{sec:Rates}

We are now concerned with estimating the expected microlensing event rate for a background AGN being lensed by a stellar-mass object in M31. We use a similar approach to \citet{Meusinger2010} in order to estimate these rates but update some parameters to reflect more recent data. These parameters are noted in Table \ref{tab:GeoParams}. Deriving a rate estimate requires an estimate both of the number density of background AGN and of the expected microlensing optical depth, i.e. the probability that a background source falls within the projected Einstein radius of a lensing object.

\begin{table}Geometric Parameters
    \centering
    \begin{tabular}{c|r|l}
    M31 ``redshift'' & $1.775\times10^{-4}$ & \\
    M31 distance$^1$ & 785 & kpc \\
    M31 inclination$^1$ & 77.5$^\circ$ & \\
    M31 PA$^2$ & 37$^\circ$ & \\
    \hline
    S21 redshift$^3$ & 2.109 \\
    S21 separation from M31 & $26.2\arcmin$ & \\
    S21 projected separation & 6.0 & kpc \\
    \hline
    Critical denisty: & & \\
    $D_{\rm{s}}$ & 1758.58 & Mpc \\
    $D_{\rm{d}}$ & 0.785 & Mpc \\
    $D_{\rm{ds}}$ & 1758.33 & Mpc \\
    $\Sigma_{\rm crit}$ & $4.4\times10^{3}$ & $\rm{kg\,m^{-2}}$ \\
    & $2.1\times10^{6}$ & $\rm{M_\odot\,pc^{-2}}$ \\
    & $3.1\times10^{7}$ & $\rm{M_\odot\,arcsec^{-2}}$ \\
    \hline
    Quasar counts$^4$, $n_{\rm QSO}$: \\
    15.5 < g < 21 & 64  & $\rm{deg^{-2}}$ \\
    15.5 < g < 22 & 141  & $\rm{deg^{-2}}$ \\
    15.5 < g < 23 & 271 & $\rm{deg^{-2}}$ \\
    15.5 < g < 24 & 487 & $\rm{deg^{-2}}$ \\
    15.5 < g < 25 & 840 & $\rm{deg^{-2}}$ \\ 
    \end{tabular}
    \caption{Parameters used in the geometric event rate analysis. The ``redshift'' for M31 corresponds to the value required to obtain an angular diameter distance of 785 kpc. [1]\citet{Geehan2006}; [2]\citet{Tamm2012}; [3]\citet{Meusinger2010}; [4]\citet{Palanque2016}.}
    \label{tab:GeoParams}
\end{table}

The optical depth to microlensing can be obtained via $\tau=\Sigma_{*}/\Sigma_{\rm cr}$, where $\Sigma_{*}$ is the stellar surface-mass density and $\Sigma_{\rm cr}$ is the critical surface-mass density given by:

\begin{equation}
\Sigma_{\rm cr}=\frac{c^2}{4\pi G}\frac{D_{\rm s}}{D_{\rm d}D_{\rm ds}}
\end{equation}

Where $D_{\rm s}, D_{\rm d}, D_{\rm ds}$ are the angular diameter distances to the source, lens and lens--source distance respectively. The critical surface density is noted in Table \ref{tab:GeoParams}. With the optical depth estimates in hand the expected event rate can be found using:

\begin{equation}
\Gamma=\frac{2N_{\rm QSO}\tau}{\pi t_{\rm E}}
\end{equation}

For these rate calculations we make the simplifying assumption that the event timescale is the same for all events and set $t_{\rm E}=1\,{\rm yr}$, appropriate to the Sharov21 event.

In order to determine the total number of background QSOs we make use of the estimates reported in \citet{Palanque2016}. The expected number densities for various apparent magnitude ranges are noted in Table \ref{tab:GeoParams}. One further factor to consider is the ability for any survey to reliably detect AGN through the disc of M31. \citet{Dalcanton2012} note that typical extinction values in M31 are on the order of $A_{\rm V}\sim1$ with less than 10\% of sightlines displaying an $A_{\rm V} > 2$. The \citet{Vilardell2006} survey reports a limiting magnitude of 25.5 and 26.0 in $V$ and $B$ respectively though we note that our candidate events are all brighter than $B\sim25$ at their faintest epoch. In our rate estimates we therefore assume that the 15.5 < g < 24 QSO density is the most appropriate choice for our rate calculations and allows for a reliable detection of the full light curve for any event.

What remains is for us to derive an estimate of the expected stellar surface mass density at a given distance from the M31 centre. For this we make use of the stellar mass models from \citet{Tamm2012} and for simplicity focus on their two-component model. This model consists of both a bulge and disc ellipsoid, each with rotational symmetry, to describe the stellar-mass profile of M31. When projected onto the sky, these components provide us with stellar surface-mass densities along the major and minor axes of M31. These are shown in Figures \ref{fig:TammMajor} and \ref{fig:TammMinor} and have been truncated to $3\leq R_{\rm proj}\leq30,000\,{\rm pc}$. Armed with this information, we are now in a position to define isodensity elliptical contours with corresponding semi-major and semi-minor distances. This in turn allows us to define elliptical annuli of assumed constant optical depth with which to derive our rate estimates across the extent of M31.

\begin{figure}
	\includegraphics[width=\columnwidth]{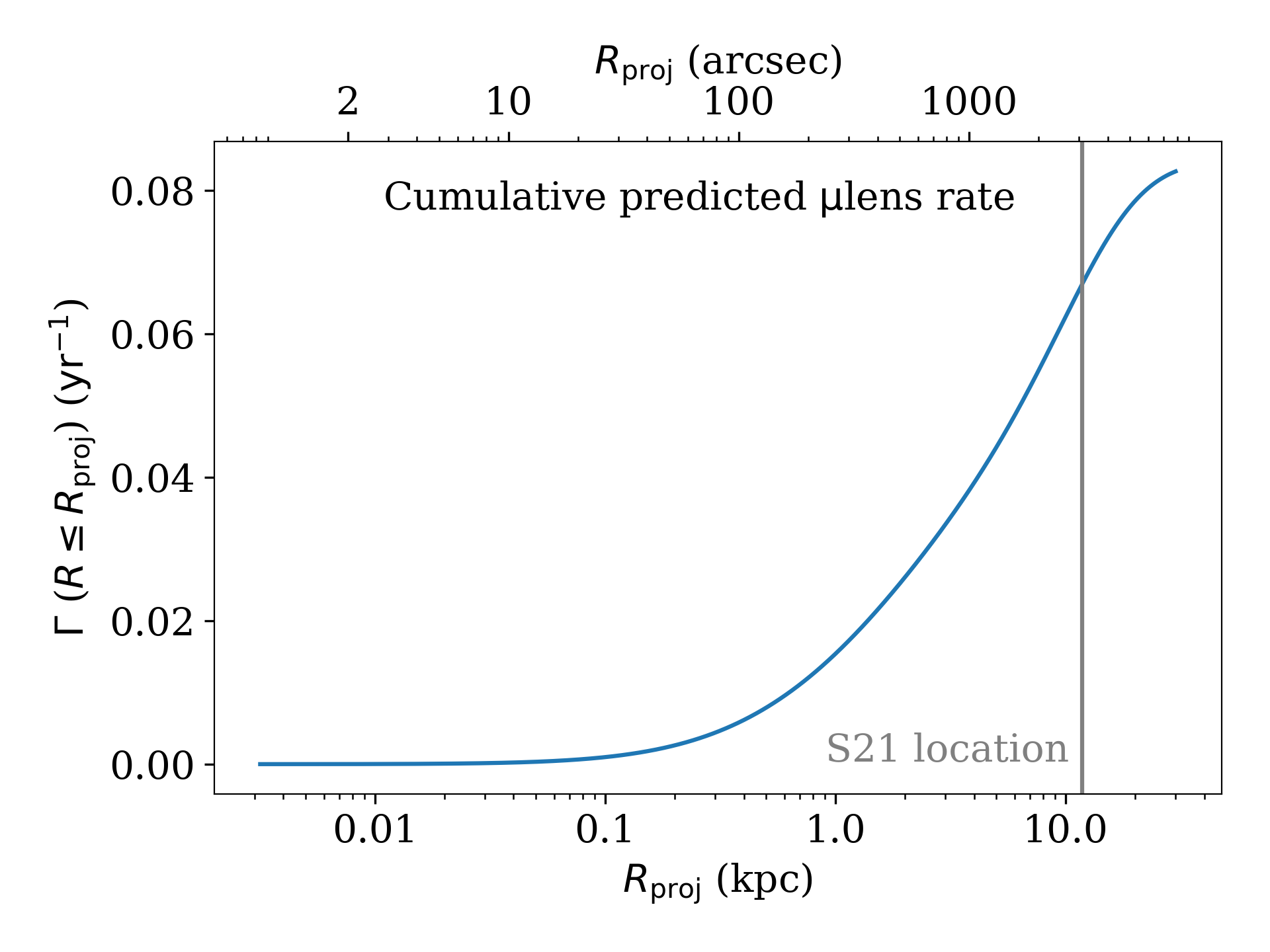}
    \caption{Cumulative predicted microlensing rate against the M31 major axis. The rates are calculated using isodensity ellipses based on the two-component mass model described in \citet{Tamm2012}. The vertical line shows the location of the isodensity ellipse appropriate to Sharov21.}
    \label{fig:microRate}
\end{figure}

Figure \ref{fig:microRate} shows the cumulative microlensing rate as a function of distance along the M31 major axis. We estimate that the overall event rate for any background AGN to fall within the Einstein radius of a stellar lens in M31 to be $\Gamma=0.0826\,{\rm yr}^{-1}$ giving an average timescale between events of $\sim12\,{\rm yr}$. We can add a further geometric constraint to this estimate if we require a higher amplitude event. The peak magnification for a point-source, point lens microlensing event is determined by the impact parameter and at $y_0=1$ this corresponds to a magnification $\mu\simeq1.34$. A minimum magnification factor of two, or thirty in the case of Sharov21, requires $y_0\leq0.556$ or $y_0\leq0.033$ respectively. The areas subtended by these higher magnification regions are reduced by the square of these values giving rise to an average timescale between events of this type of approximately 40 and 11,000 years respectively. These simplified estimates confirm the exceptional nature of the Sharov21 event but also allow for the possibility of detecting a number of lower amplitude events on more reasonable timescales.

One point to note is that though the optical depth to microlensing is greatest at the innermost radii, the area subtended by these regions is considerably smaller than for more distant regions with lower optical depth and thus contribute only a small fraction to the overall rate estimates. The optical depth falls within the range $5.6\times10^{-2}\leq \tau\leq1.3\times10^{-6}$. With these values we are still safely in the low optical depth regime. At the location of Sharov21 we estimate $\tau\simeq6\times10^{-5}$ and $M_{\rm surface}\simeq135\,M_\odot\,{\rm pc}^{-2}$. We have plotted the isodensity ellipse corresponding to the position of Sharov21 in Figure \ref{fig:M31}, with a semi-major axis distance of approximately 12 kpc or $3100\arcsec$. Given that our lensing rate is proportional to the product of the surface-mass density and the area subtended by any isodensity annulus, there is a turnover in the derived rates, centred at approximately 5 kpc or $1300\arcsec$ along the major axis. This location, where the microlensing rate effectively peaks, is highlighted by the inner annulus in Figure \ref{fig:M31}.

\subsection{Event Rate Discussion}

Our event rate estimate is likely an oversimplification but it suggests that the average time between microlensing events of this nature, with a factor of two or greater amplification, is on the order of half a century and would occur most frequently at intermediate radii from the M31 centre. In these calculations, given the close proximity of M31 to us as observers, the optical depth is relatively insensitive to the redshift of the source though the assumption of a point-source will break down for some lens geometries. In these cases the peak amplification of the event will be lower and there may be evidence for chromatic changes as a consequence of the AGN disc being partially resolved  during the event. We would also expect that, assuming a binary fraction of $\sim50\%$, approximately 10\% of events would show additional structure in their light curve due to the presence of a binary lens where a favourable alignment produces notable deviations from the symmetric point-lens case. One additional factor to consider is that the blending of the target flux with other M31 sources and/or a strong host galaxy component has the possibility of further diluting any microlensing signal. The number of confirmed AGN behind M31 remains low at present \citep{Massey2019} though this is certain to evolve in the coming years. It seems clear that spectroscopic confirmation will be required of any candidate to confirm their status as a background AGN. It is not trivial to perform a colour selection given that many of these background AGN will be seen through a not-insignificant dust column.

Given the low probability of observing a background AGN microlensing event it is thus perhaps surprising that we have been able to identify two other candidate AGN microlensing events in the \citet{Vilardell2006} data. Both of these appear within a 4 year timeframe and $\simeq0.3\,\rm{deg}^2$ survey footprint. T2 is spectroscopically confirmed as a background AGN and T3, as an X-ray source, remains a likely candidate for a background AGN. The peak amplifications for these additional events are a factor 2--3. Though much less than the factor 30 seen in Sharov21, these low-amplitude microlensing events will be far more likely to occur in general. We must therefore allow for the possibility that, either our hypothesis is incorrect, or that there may be an additional population of stellar-mass lenses which are as yet unaccounted for.

\subsubsection{Non-stellar microlenses?}

Current cosmological \citep[e.g.,][]{Planck2018} and galactic dynamics inform us that only 20\% of the observed mass density is in baryonic form. One candidate for this non-baryonic (`dark') matter are massive astrophysical compact halo objects \citep[MACHOs;][]{Paczynski1986}. MACHOs could be a range of non-luminous astrophysical objects,  but would readily reveal themselves via gravitational microlensing events. With the recent detection of gravitational waves from merging black holes \citep{Abbott2016, Abbott2019} the interest in primordial black holes (PBHs), which are potential MACHO candidates, has been reinvigorated \citep[e.g., ][]{Clesse2017, Stegmann2019}. The MACHO experiment was designed to detect the microlensing of stars in the Magellanic Clouds by compact bodies in the Milky Way Galactic halo \citet{Alcock1996}.  Using the results from the MACHO collaboration \citep{Alcock2000}, and a range of Galactic models for LMC microlensing, \citet{Hawkins2011} finds that the a MACHO contribution to the MW halo is not ruled out, and the MACHO content could potentially be around 20\%. Recent analyses from \citet{Calcino2018} using microlensing constraints towards the Large Magellanic Cloud suggest that although the likelihood for $\sim$1-10 M$_{\odot}$ objects is weakened, the constraints for masses around 10 M$_{\odot}$ are still viable.

As such, though we generally remain agnostic, we acknowledge the potential existence for MACHOs in the M31/MW halo and note that the presence of a population of these objects may help to explain the discrepancy between our predicted and observed microlensing rates. Suffice to say, this provides an additional and compelling reason to undertake a systematic search for more events of this nature and on these year-long timescales. Of particular import will be the positions of the events relative to the M31 centre. Our rate analysis indicates that the most favourable location for detecting microlensing events is located nearer to the centre of M31 than the positions of our three current microlensing candidates. It is possible that the blending effects described above are simply biasing us towards detecting events at greater radii. It is also possible that the outskirts of the stellar disc of M31 play host to an additional repository of lenses as yet unaccounted for. Further speculation at this stage is premature as we are currently limited by a low number of candidate events.

\section{Conclusions}

We have re-examined the decades-long lightcurve for the object known as Sharov21, a background AGN within M31 seen to undergo a thirty-fold increase in brightness over one year. Armed with only the lightcurve, a point-source/point-lens microlensing model and assumption of a stellar-mass lens it has been possible to derive constraints on the expected distance of the lens which is in excellent agreement with the true distance of M31. We believe this provides strong evidence that Sharov21 was indeed a rare, high-amplitude microlensing event. We find that slight discrepancies with the data are not sufficient, given the data quality, to justify discarding the simple model in favour of more complex lensing scenarios.

We have analysed archival data on Sharov21 from multiple sources and our resulting SED shows that this AGN can be considered an otherwise unremarkable type-I AGN. The high resolution Hubble data from the PHAT survey confirms that Sharov21 is consistent with a point source.

In addition to our work on Sharov21 we have undertaken a search for additional microlensing candidates that display characteristics of a simple microlensing event on similar timescales in the a four-year survey of a sector of M31. This search has yielded 20 candidate events, one is a confirmed background AGN and the other an X-ray source and thus a promising background candidate. For the confirmed background AGN, our microlensing analysis shows that this event is also consistent with the lens object being located at the distance of M31.

Our exploration of the expected microlensing event rate shows that these events should occur on average every half century or so. This is a higher rate than derived in \citet{Meusinger2010} but not high enough to explain our current number of candidates if our microlensing hypothesis is correct. This may suggest the presence of an additional population of lensing objects in the outskirts of M31 which is not yet accounted for.

A detailed, systematic search for long-term microlensing candidates in M31, including spectroscopic follow-up, is required in order to address the discrepancy between our observed and predicted rates. The timescales for these events are on the order of years and are most likely to occur toward intermediate radii from the M31 centre. These events can in principle provide valuable information about these distant AGN if the data is well sampled. Perhaps more importantly, by monitoring as many background sources as possible it allows a probe of the M31 stellar and dark halo populations in both M31 and the Milky Way. The timescales for these events are longer and stand in contrast to the microlensing events of source stars in the LMC/SMC/M31 by intervening compact objects. A regime worthy of further exploration.

\section*{Acknowledgements}

The authors would also like to thank H. Meusinger for providing the Sharov21 light curve data used in their paper.

NPR acknowledges support from the STFC and the Ernest Rutherford Fellowship scheme.

The Pan-STARRS1 Surveys (PS1) and the PS1 public science archive have been made possible through contributions by the Institute for Astronomy, the University of Hawaii, the Pan-STARRS Project Office, the Max-Planck Society and its participating institutes, the Max Planck Institute for Astronomy, Heidelberg and the Max Planck Institute for Extraterrestrial Physics, Garching, The Johns Hopkins University, Durham University, the University of Edinburgh, the Queen's University Belfast, the Harvard-Smithsonian Center for Astrophysics, the Las Cumbres Observatory Global Telescope Network Incorporated, the National Central University of Taiwan, the Space Telescope Science Institute, the National Aeronautics and Space Administration under Grant No. NNX08AR22G issued through the Planetary Science Division of the NASA Science Mission Directorate, the National Science Foundation Grant No. AST-1238877, the University of Maryland, Eotvos Lorand University (ELTE), the Los Alamos National Laboratory, and the Gordon and Betty Moore Foundation.




\bibliographystyle{mnras}
\bibliography{Bruce}




\appendix

\section{Supplementary figures}

\begin{figure*}
	\includegraphics[width=\columnwidth*2]{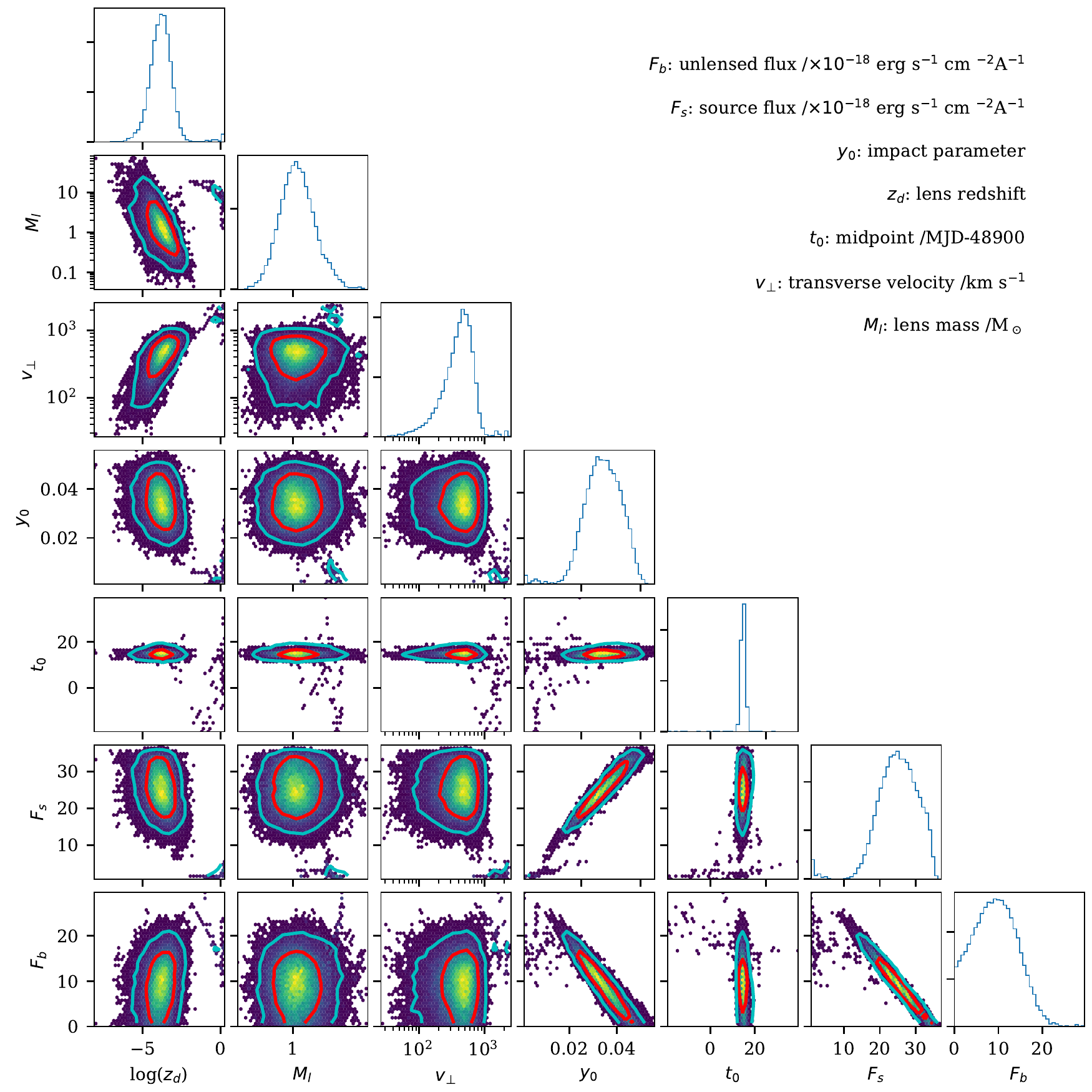}
    \caption{Full corner plot for the MCMC microlensing analysis for Sharov21.}
    \label{fig:S21_corner}
\end{figure*}

\begin{figure*}
	\includegraphics[width=\columnwidth*2]{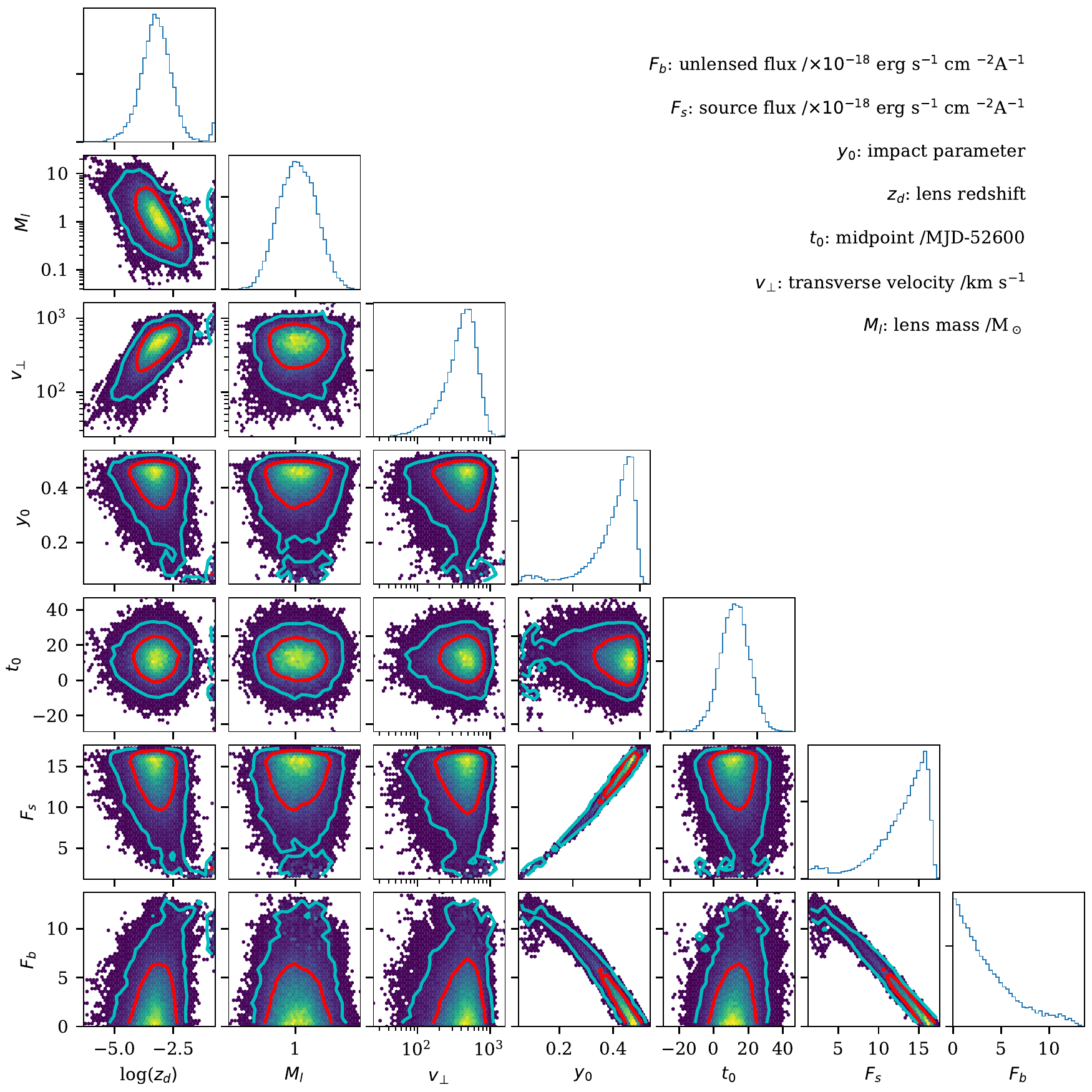}
    \caption{Full corner plot for the MCMC microlensing analysis for candidate T2.}
    \label{fig:T2_corner}
\end{figure*}

\begin{figure}
	\includegraphics[width=\columnwidth]{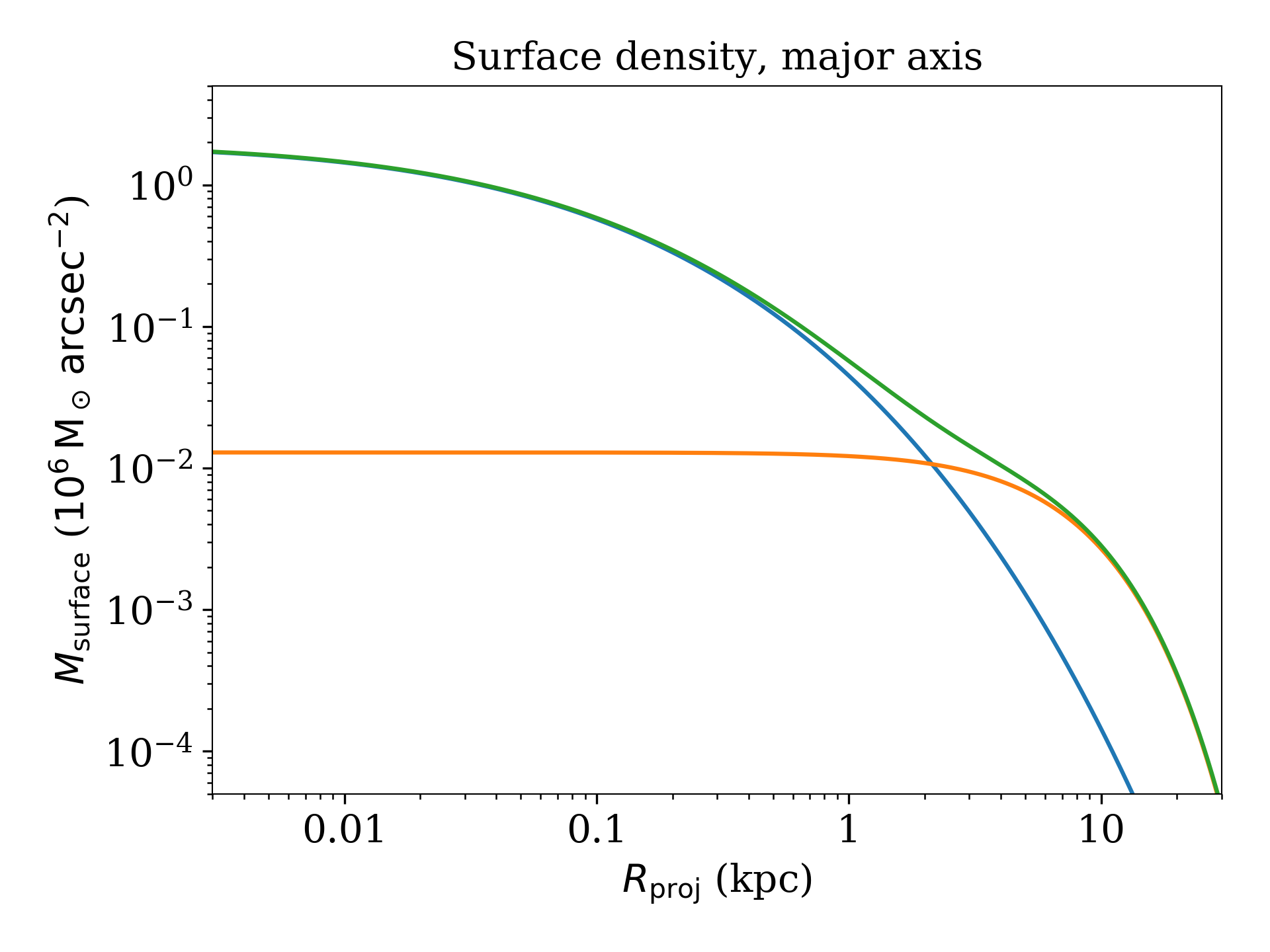}
    \caption{M31 stellar surface density along the major axis. Derived from the two-component model described in \citet{Tamm2012}.}
    \label{fig:TammMajor}
\end{figure}

\begin{figure}
	\includegraphics[width=\columnwidth]{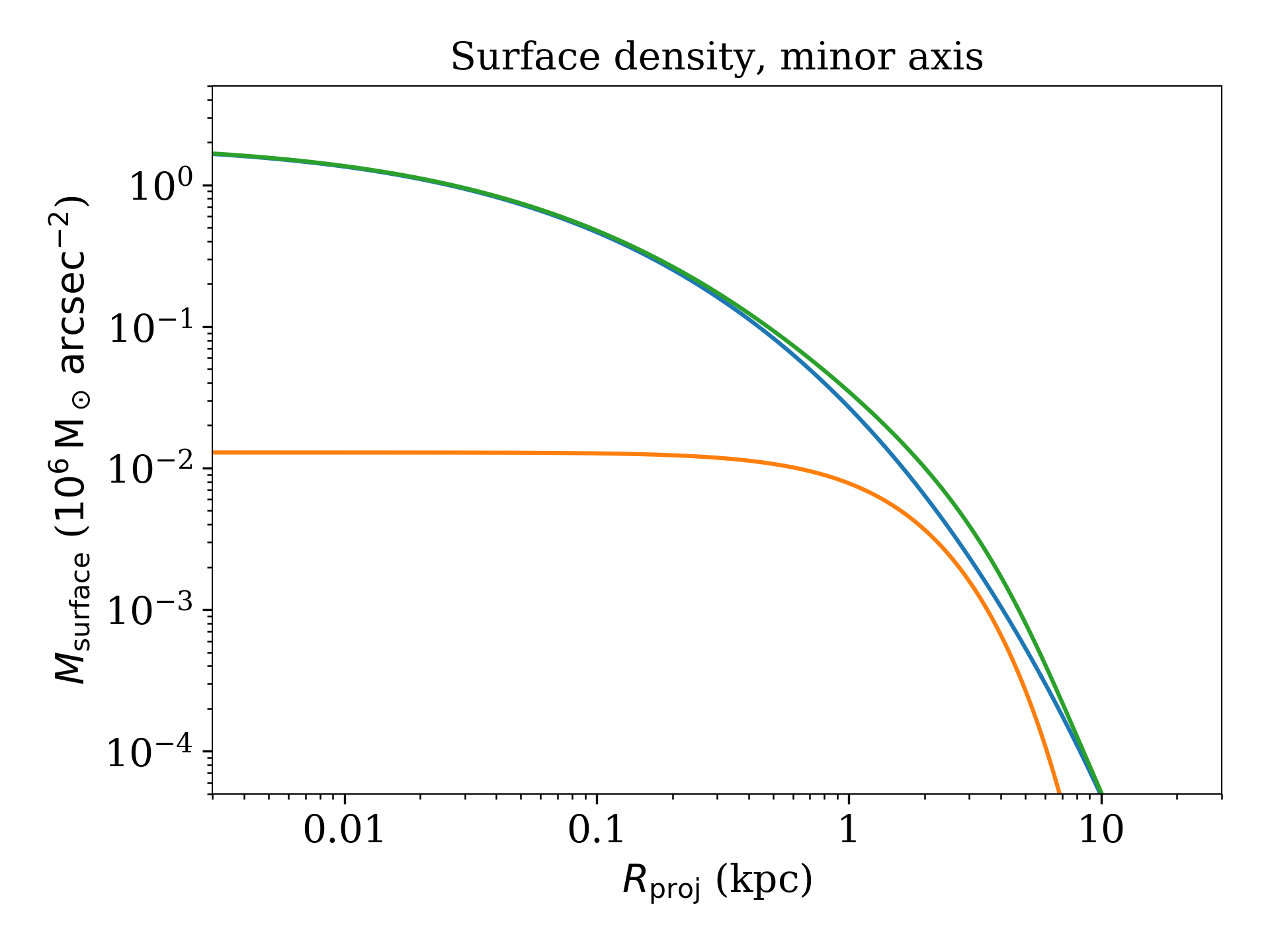}
    \caption{M31 stellar surface density along the minor axis. Derived from the two-component model described in \citet{Tamm2012}.}
    \label{fig:TammMinor}
\end{figure}


\bsp	
\label{lastpage}
\end{document}